\documentclass{article}

\begin{document}

\title{A slowly rotating perfect fluid body in an ambient vacuum}
\author{Michael Bradley$^1$, Gyula Fodor$^{2,3}$
and Zolt{\'a}n Perj{\'e}s$^2$ \\
$1$ Department of Plasma Physics, Ume{\aa} University, S--901 87\\
Ume{\aa}, Sweden \\
$2$ KFKI Research Institute for Particle and Nuclear Physics,\\
H--1525, Budapest 114, P.O.B.\ 49, Hungary \\
$3$ \ \  Department of Physics, Waseda University, 3-4-1 Okubo, \\
Shinjuku, Tokyo 169-8555, Japan}
\date{\today }
\maketitle

\begin{abstract}
A global model of a slowly rotating perfect fluid ball in general relativity
is presented. To second order in the rotation parameter, the junction
surface is an ellipsoidal cylinder. The interior is given by a limiting case
of the Wahlquist solution, and the vacuum region is \emph{not}
asymptotically flat. The impossibility of joining an asymptotically flat
vacuum region has been shown in a preceding work.
\end{abstract}

\section{Introduction}

In a preceding paper \cite{bfmp}, henceforth called \textbf{I}, the
impossibility of matching the Wahlquist metric to an asymptotically flat
vacuum domain was shown. This result is not too surprising in the light of
investigations in \cite{MS}, where the tendency of the matching conditions
to be overdetermined has been pointed out. However, it would be very
embarrasing from the point of view of general relativity if this matching
turned out to be impossible to any vacuum region. In this paper, the problem
of matching of the slowly rotating Wahlquist metric to a more general vacuum
exterior is investigated to a precision of quadratic order in the angular
velocity.

An approximation scheme for slow rotation was first introduced by Brill and
Cohen \cite{BriCoh,CohBri} to first order in the small angular velocity. In
our paper we are using the the formalism developed by Hartle\cite{Hart},
taking into account quadratic-order terms in the power series expansion in
the angular velocity $\Omega $ of the fluid. The metric of both
regions has the form \newpage
\begin{eqnarray}
ds^2 &=&(1+2h)A^2dt^2-(1+2m)B^2dr^2  \nonumber \\
&&-(1+2k)C^2\left\{ d\vartheta ^2+\sin ^2\vartheta \left[ d\varphi +\left(
\Omega -\omega \right) dt\right] ^2\right\}   \label{ds2}
\end{eqnarray}
where the static, spherically symmetric state is described by the functions $%
A$, $B$ and $C$ depending only on the radial coordinate $r$, while the
functions $\omega $, $h$, $m$ and $k$ can, in general, depend both
on $r$ and $\vartheta $. The rotation potential $\omega $ is of first order
in the angular velocity $\Omega $, and the functions $h$, $m$ and $k$ are of
order $\Omega ^2$.

In the Wahlquist interior domain, the rotation potential $\omega $ is a
function of the radial coordinate alone. Hence, from the juction conditions
it follows, that $\omega $ is independent of $\vartheta $ in the vacuum
region as well. 
In our model of the space-time we drop the condition of asymptotic flatness,
and we perform the matching with the most general vacuum metric with $\omega
=\omega (r)$. Likewise the expansions of the second-order metric functions
in Legendre polynomials are sought in the form $h=h_0+h_2P_2(\cos \vartheta )
$, $m=m_0+m_2P_2(\cos \vartheta )$, $k=k_2P_2(\cos \vartheta )$, 
where the functions $h_0$, $h_2$, $m_0$, $m_2$ and $k_2$ depend only on the
radial coordinate $r$, and $P_2(\cos \vartheta )=(3\cos ^2\vartheta
-1)/2$. 

In Sec. 2 of this paper, we present the form of these functions for the
vacuum domain, and investigate the effect of the non asymptotically flat
part of the perturbations on the curvature. In Sec. 3., the junction
conditions are calculated, and the constants determining the exterior metric
as functions of the unperturbed radius $r_1$ are given in full detail.

\section{The Vacuum Exterior}

In this section we consider the form of the vacuum metric to the required
accuracy. The unperturbed metric is described by the Schwarzschild solution, 
$A^2=1/B^2=1-2M/r$ and $C=r$. The solution for the perturbed metric is of
the form (\ref{ds2}), with $\Omega =0$ and 
\begin{equation}
\omega =\frac{2aM}{r^3}\ ,
\end{equation}
where the additive constant of integration has been removed by a rigid
rotation. Integration of the second-order metric functions yields 
\begin{eqnarray}
h_0 &=&\frac 1{r-2M}\left( \frac{a^2M^2}{r^3}+\frac r{2M}c_2\right) \\
m_0 &=&\frac 1{2M-r}\left( \frac{a^2M^2}{r^3}+c_2\right) \\
h_2 &=&3c_1r\left( 2M-r\right) \log \left( 1-\frac{2M}r\right) +a^2\frac
M{r^4}\left( M+r\right)  \nonumber \\
&&+2c_1\frac Mr\left( 3r^2-6Mr-2M^2\right) \frac{r-M}{2M-r}+\left( 1-\frac{2M%
}r\right) r^2q_1  \label{eqh2} \\
k_2 &=&3c_1(r^2-2M^2)\log \left( 1-\frac{2M}r\right) -a^2\frac M{r^4}(2M+r) 
\nonumber \\
&&-2c_1\frac Mr(2M^2-3Mr-3r^2)+\left( 2M^2-r^2\right) q_1 \\
m_2 &=&6a^2\frac{M^2}{r^4}-h_2 \ .  \label{eqk2}
\end{eqnarray}
In this approximation, the slowly rotating solution is characterized by the
mass $M$, the first order small rotation parameter $a$, and the second order
small constants $c_1$, $c_2$ and $q_1$.

With the special choice $q_1=0$, the metric is asymptotically flat, and this
form was used in \cite{bfmp} to show the impossibility of matching the
Wahlquist solution to an asymptotically flat vacuum. Since the asymptotic
behaviour changes completely when $q_1$ becomes nonzero, the far field
behaviour cannot be treated in a perturbative way, and the present series
expansion is certain to hold only within an open neighborhood of the
junction surface. This is also indicated by the fact that the terms
containing the integration constant $q_1$ tend to $r^2$ as $r\to \infty $.

To explore the effects of the $q_{1}$ perturbations on the curvature, we use
a canonical locally nonrotating Lorentz tetrad for the metric (\ref{ds2}): 
\begin{eqnarray}
e_0&=&\left(1-h+\frac{r^2}{2A^2}\omega^2 \sin^2\vartheta\right) \frac{1}{A}%
\frac{\partial}{\partial t} \\
e_1&=&(1-m) A\frac{\partial}{\partial r} \\
e_2&=&(1-k) r^{-1}\frac{\partial}{\partial\vartheta} \\
e_3&=&-\frac{r}{A^2} \omega\sin\vartheta \frac{\partial}{\partial t}
-\left(1-k+\frac{r^{2}}{2A^{2}}\omega^2 \sin^2\vartheta\right) \frac{1}{%
r\sin\vartheta}\frac{\partial}{\partial\varphi}\ .
\end{eqnarray}

In this tetrad, the gravitoelectric ($E_i$) and gravitomagnetic ($H_i$)
fields are defined in terms of the Riemann tensor as follows \cite{fmp}: 
\begin{eqnarray}
E_1 &=&R_{0101}\qquad E_2=R_{0202}\qquad \ \ E_3=R_{0102} \\
H_1 &=&R_{0123}\qquad H_2=-R_{0213}\qquad H_3=R_{0223}\ .
\end{eqnarray}
In the asymptotic region, $r\to \infty $, the gravimagnetic part of the
curvature goes to zero. The gravielectric components tend to the finite
values 
\begin{eqnarray}
\lim_{r\to \infty }E_1 &=&\left( 1-3\cos ^2\vartheta \right) q_1  \label{E_1}
\\
\lim_{r\to \infty }E_2 &=&\left( 1-3\sin ^2\vartheta \right) q_1 \\
\lim_{r\to \infty }E_3 &=&3\sin \vartheta \cos \vartheta\ q_1\ .  \label{E_3}
\end{eqnarray}
The algebraic structure of the Weyl tensor at infinity, (\ref{E_1})-(\ref
{E_3}), guarantees that the components cannot all be transformed to zero,
and hence the space-time cannot be asymptotically flat. However, these
values are obtained by means of perturbative calculations, and higher-order
corrections may contribute by divergent terms\footnote{%
We remind the reader that the asymptotic region may lie outside the
domain of convergence of the power series expansion in the angular velocity.}
to the limiting values (\ref{E_1})-(\ref{E_3}).

The angular behavior of the gravielectric field indicates that a quadrupolar
mass distribution at large distance may act as the source of the deviations
from asymptotic flatness. To show this, let us consider a large sphere with
radius $R$, and with a non-uniform surface density distribution 
\begin{equation}
\mu =\sum_{i=0}^\infty \mu _{2i}P_{2i}(\cos \vartheta )\ ,
\end{equation}
where $\mu _{2i}$ are constants. In the weak-field approximation,
linearizing around the background Minkowski metric and expanding
in powers of $1/R$, the metric near the center of the sphere can be written
as 
\begin{equation}
ds^2=(1+2\psi )dt^2-(1-2\psi )(dr^2+r^2d\vartheta ^2+r^2\sin ^2\vartheta
d\varphi ^2)
\end{equation}
where 
\begin{equation}
\psi =-\frac{1}{2} \mu _0R-\frac 1{10R} \mu _2r^2P_2(\cos \vartheta )+O\left(
\frac 1{R^3}\right) \ .
\end{equation}
(With the conventions used in \textbf{I} the gravitational constant is $\frac{1}{8\pi}$.)
The $\mu _0$ terms can be absorbed by rescaling the $t$ and $r$ coordinates,
while the $\mu _2$ terms have exactly the same angular and radial dependence
as the $q_1$ terms in the far field exterior vacuum region in (\ref{eqh2})-(%
\ref{eqk2}). The higher than quadrupole mass distribution coefficients can
be arbitrary, which shows that they cannot be determined from a slow
rotation formalism which takes into account only up to quadratic order terms
in the angular velocity $\Omega $. Instead of using the linearized gravity
approximation, an exact static but not spherically symmetric 
Weyl-class vacuum metric could also be constructed. But since for
slow enough rotation there is always a region where the $r^2q_1$ terms are
small whereas the other terms in the metric are negligible, no new
insight would result from that analysis. 

In a general stationary axisymmetric vacuum exterior solution the suitable
hypersurfaces for matching to an interior fluid region are determined in the
paper of Roos\cite{Roos}. In the limit of no rotation, the matching
surface is the history of the sphere $r=r_1$. For slow rotation the surface
becomes an ellipsoid, and its deformation is described by 
\begin{equation}
r=r_1+a^2\left[ \chi _0+\chi _2P_2(\cos \vartheta )\right]   \label{SV}
\end{equation}
where $a$ is the small rotational parameter and $\chi _0$ and $\chi _2$ are
constants to be determined by the matching conditions. The expressions for
the normal vector and the extrinsic curvature of the matching surface have
the same forms as in \textbf{I}.

\section{Matching with the Wahlquist solution}

In \textbf{I} we have computed the second order form of the Wahlquist metric 
\cite{Wahl} in Hartle's coordinates. The metric functions and the
zero-pressure matching surface is expressed in I in terms of the constants $%
\mu _0$ and $\kappa $ characterizing the static configuration and in terms
of the the small parameter $r_0$ which is proportional to the angular
velocity. These results will be used now for matching with the
vacuum solution of the previous section. The radial coordinate in the fluid
region is denoted by $x$ instead of $r$. In the limit of no rotation the
fluid region is described by the Whittaker metric \cite{Whitt}, and the
matching surface is the sphere characterized by $x=x_1$, where $x_1\cot
x_1=\kappa ^2$ (\textit{cf.} I).

For slow rotation the matching surface $\mathcal{S}$ is an ellipsoidal
cylinder characterized by a vanishing pressure and by the embedding
condition (\ref{SV}). We equate with each other the respective induced
extrinsic curvatures $K_{(V)}$ and $K_{(W)}$ of $\mathcal{S}$, in the vacuum
and in the Wahlquist region. Hence, the equations of matching are 
\begin{equation}
ds_{(V)}^2|_{\mathcal{S}}=ds_{(W)}^2|_{\mathcal{S}}\qquad K_{(V)}|_{\mathcal{%
S}}=K_{(W)}|_{\mathcal{S}}\ .  \label{matchcd}
\end{equation}
where $ds^2|_{\mathcal{S}}$ is the induced metric. The values of the metric
coefficients and their derivatives on $\mathcal{S}$ are given by a power
series expansion in $r_0$ in the fluid and in $a$ in the vacuum regions,
respectively. 
We apply a rigid rotation in the fluid region by setting $\varphi \to
\varphi +\Omega t$ where $\Omega $ is a constant. Then we re-scale the
interior time coordinate $t$ by 
\begin{equation}
t\to c_4(1+r_0^2c_3)t  \label{tscale}
\end{equation}
with further constants $c_3$ and $c_4$ to be determined from the matching
conditions.

Substituting in the matching conditions (\ref{matchcd}), and Taylor
expanding to second order in the angular velocity, we get a set of
linear equations for the parameters $\chi _0$, $\chi _2$, $c_1$, $c_2$, $c_3$
and $q_1$ . Here we list the solution of the matching equations for the
constant $c_3$ of the interior time scaling and for the constants $q_1$, $c_1
$, $c_2$, $\chi _0$ and $\chi _2$ of the vacuum domain in terms of the
radius $x_1$ of the Whittaker fluid ball, the rotation parameter $a$ given
by 
\begin{eqnarray}
a&=&\frac{r_0}{3\cos x_1}\, \frac{2x_1\cos^2x_1-3\sin x_1\cos x_1+x_1}{\sin
x_1\cos x_1-x_1}  \label{a}
\end{eqnarray}
and the density $\mu _0$. The results of \textbf{I} for the matching to zero
order in the angular velocity 
\begin{eqnarray}
\Omega&=&\frac{\mu_0x_1 r_0}{6\sin x_1\cos x_1} \\
\end{eqnarray}
hold without any change, 
\begin{eqnarray}
M&=&\frac{r_1}{2\kappa^2}(\kappa^2-\cos^2x_1) \\
r_1&=&\frac{2^{1/2}}{\kappa{\mu_0}^{1/2}}\sin x_1 \\
c_4&=&\cos x_1 \ .
\end{eqnarray}
The second-order matching conditions have the solution 
\begin{eqnarray}
c_3 &=&-\frac{\cos {x_1}\mu _0}{72l_3^2\cos ^2{x_1}\sin {x_1}}[(15\cos ^2{x_1%
}+7\sin ^2{x_1})x_1^3  \nonumber \\
&-&(18\cos ^2{x_1}+4\sin ^2{x_1})x_1^2\sin {x_1}\cos {x_1}+12\cos ^3{x_1}%
\sin ^3{x_1} \\
&-&3x_1\cos ^2{x_1}\sin ^4{x_1}-9x_1\cos ^4{x_1}\sin ^2{x_1}]  \nonumber \\
q_1 &=&\frac{a^2\cos ^2{x_1}\mu _0^2x_1^3}{2l_2^2l_3^3\sin ^3{x_1}} 
\nonumber \\
&\times &[51x_1^3\cos ^3{x_1}-27x_1^2\cos ^4{x_1}\sin {x_1}-27x_1\cos ^5{x_1}%
\sin ^2{x_1}  \nonumber \\
&+&3\cos ^4{x_1}\sin ^3{x_1}-24x_1\cos ^3{x_1}\sin ^4{x_1}  \nonumber \\
&-&34x_1^2\cos ^2{x_1}\sin ^3{x_1}+8x_1^3\cos {x_1}\sin ^2{x_1}-x_1^4\sin {%
x_1}]  \nonumber \\
&+&\frac{3a^2l_1\cos ^5{x_1}\mu _0^2x_1^5}{l_2^2l_3^4\sin ^3{x_1}}\log
\left( \sin {x_1}\cos {x_1}/x_1\right) \\
c_1 &=&\frac{a^2l_1\cos ^5{x_1}\mu _0^2x_1^5}{l_2^2l_3^4\sin ^3{x_1}} \\
c_2 &=&\frac{-a^2l_3\cos ^{1/2}{x_1}\sin ^{1/2}{x_1}\mu _0^{1/2}}{%
2^{5/2}l_2^2x_1^{3/2}}  \nonumber \\
&\times &[9x_1^3\cos ^2{x_1}-27x_1\cos ^4{x_1}\sin ^2{x_1} \\
&+&18\cos ^3{x_1}\sin ^3{x_1}-9x_1\cos ^2{x_1}\sin ^4{x_1}+2x_1^2\cos {x_1}%
\sin ^3{x_1}+x_1^3\sin ^2{x_1}]  \nonumber \\
\chi _0 &=&\frac{3l_3\cos ^{3/2}{x_1}\mu _0^{1/2}}{2^{3/2}x_1^{1/2}l_2^2\sin
^{3/2}{x_1}}(x_1^2\cos {x_1}-2x_1\cos ^4{x_1}\sin {x_1} \\
&+&\cos ^3{x_1}\sin ^2{x_1}+x_1\sin ^3{x_1}-\cos {x_1}\sin ^4{x_1}) 
\nonumber \\
\chi _2 &=&\frac{-3l_3\cos ^{3/2}{x_1}\mu _0^{1/2}}{2^{3/2}x_1^{1/2}l_2^2%
\sin ^{3/2}{x_1}x_1}(x_1^2\sin {x_1}+4x_1^3\cos ^3{x_1}-8x_1\cos ^3{x_1}\sin
^2{x_1} \\
&+&3\cos ^2{x_1}\sin ^3{x_1}-4x_1\cos {x_1}\sin ^4{x_1})  \nonumber
\end{eqnarray}
where 
\begin{eqnarray}
l_1 &=&3x_1^2+6x_1\cos {x_1}\sin {x_1}-\cos ^2{x_1}\sin ^2{x_1}-8\sin ^2{x_1}
\\
l_2 &=&x_1(3\cos ^2{x_1}+\sin ^2{x_1})-3\sin {x_1}\cos {x_1} \\
l_3 &=&x_1-\sin {x_1}\cos {x_1}
\end{eqnarray}

It is easy to show that $q_{1}$ is non-zero in the allowed range of $x_{1}$, 
$0<x_{1}<\frac{\pi}{2}$.

\section{Conclusions}

In this paper we have shown that there exist configurations where a
Wahlquist fluid ball in an ambient vacuum domain is kept in equilibrium. In
our model space-time the components of the curvature tensor tend to finite
values at infinity. A far-away quadrupole mass distribution can ensure the
required non-asymptotically flat nature of the exterior vacuum.
The solution is linearization stable in the sense that there exists a family
of exact solutions corresponding to our approximate solution, which is valid
in the fluid and in at least an open vacuum region surrounding it.
This follows from the work of Roos \cite{Roos,Roos2} showing the existence
and uniqueness in a neighbourhood of the matching surface in the general
axisymmetric and stationary case. 

\section{Acknowledgments}

This work has been partially supported by the OTKA grant T022533. M.B.
wishes to acknowledge the hospitality of the KFKI in Budapest. M.B. was
partially supported by the NFR. G. F. would like to thank the support of the
Japan Society for the Promotion of Science and the hospitality of the
Physics Department of Waseda University.


\begin{thebibliography}{99}
\bibitem{bfmp}  M. Bradley, G. Fodor, M. Marklund and Z. Perj\'{e}s, \textit{%
Class. Quantum Grav.} \textbf{17}, 351 (2000), paper \textbf{I.} In
this reference, the calculation of the ellipticity of the zero-pressure
surface is erroneous. The correct statement is that for slow rotation the
body is oblate if and only if $k_2+r_0^2\xi _2\cot x$ is negative. Evaluating
this for the Wahlquist metric, it turns out that the surface is always
prolate, indicating the action of some exterior force on the fluid.

\bibitem{MS}  M. Mars and J. M. M. Senovilla, \textit{Mod.Phys.Lett.A,} 
\textbf{13},1509 (1998)

\bibitem{BriCoh}  D. R. Brill and J. M. Cohen, \textit{Phys. Rev.} \textbf{%
143,} 1011 (1966)

\bibitem{CohBri}  J. M. Cohen and D. R. Brill, \textit{Nuovo Cimento} 
\textbf{56 B,} 209 (1968)

\bibitem{Hart}  J. B. Hartle, \textit{Astrophys. J. }\textbf{150, }1005
(1967)

\bibitem{fmp}  G. Fodor, M. Marklund and Z. Perj\'{e}s, \textit{Class.
Quantum Grav.} \textbf{16,} 453 (1999).


\bibitem{Roos}  W. Roos, \textit{Gen. Rel. Grav. }\textbf{7, }431 (1976)

\bibitem{Wahl}  H. D. Wahlquist, \textit{Phys. Rev.} \textbf{172,} 1291
(1968).

\bibitem{Whitt}  J. M. Whittaker, \textit{Proc. Roy. Soc. }\textbf{A 306,} 1
(1968)

\bibitem{Roos2}  W. Roos, \textit{Gen. Rel. Grav. }\textbf{8, } 753 (1977)
\end{thebibliography}
\end{document}